\documentclass[preprint,12pt, a4paper]{elsarticle}

\usepackage{multirow}
\usepackage{amsmath,amsfonts}
\usepackage{algorithmic}
\usepackage{graphicx}
\usepackage{textcomp}
\usepackage{xcolor}
\usepackage{amssymb}
\def\BibTeX{{\rm B\kern-.05em{\sc i\kern-.025em b}\kern-.08em
 T\kern-.1667em\lower.7ex\hbox{E}\kern-.125emX}}

\usepackage[framemethod=tikz]{mdframed} %
\usepackage{soul}
\usepackage{stfloats}

\usepackage{soul}
\usepackage{colortbl}
\usepackage{adjustbox}
\usepackage{caption}
\usepackage{booktabs} 
\usepackage{multirow}
\usepackage{tikz}
\usepackage{xcolor}
\usepackage{hyperref}
\usepackage{tabularx}
\usepackage{threeparttable}
\usepackage{multicol}
\usepackage{float}
\usepackage[most]{tcolorbox}
\usepackage{xspace}
\newcounter{num_eq}
\definecolor{purplish}{HTML}{D8D0E3}
\definecolor{purplishlight}{HTML}{EBE7F1}
\definecolor{purplishdark}{HTML}{20aee5}

\newtcolorbox[]{rqbox}[2]{
    nameref=#1,
    title=\small{#1}, 
    enhanced,
    attach boxed title to top left={yshift=-6pt, xshift=8pt},
    boxed title style={size=small,boxsep=1pt},
    colframe=purplishdark,colback=white,colbacktitle=purplishdark,
    boxsep=2pt,left=2pt,right=2pt,top=6pt,bottom=2pt,middle=2pt
}

\makeatletter
\newcommand{\responsetoreviewer}[3][black]{

\expandafter\gdef\csname #2\endcsname{#3}%
\label{rev:#2}%
{\color{black}{#3}}\color{#1}\xspace}
\makeatother

\journal{Science of Computer Programming}

\begin{document}

\begin{frontmatter}



\title{Applying Large Language Models to Issue
Classification: Revisiting with Extended Data and New Models}


\author{Gabriel Aracena}

\address{Grand Canyon University, 3300 W. Camelback Rd., Phoenix, AZ, USA}

\author{Kyle Luster}

\address{Grand Canyon University, 3300 W. Camelback Rd., Phoenix, AZ, USA}

\author{Fabio Santos}

\address{Colorado State University, 456 University Ave., Fort Collins, CO, USA}

\author{Igor Steinmacher}

\address{Northern Arizona University, 1900 S. Knoles Dr., Flagstaff, AZ, USA}

\author{Marco A. Gerosa}

\address{Northern Arizona University, 1900 S. Knoles Dr., Flagstaff, AZ, USA}

\begin{abstract}
Effective prioritization of issue reports in software engineering helps to optimize resource allocation and information recovery. However, manual issue classification is laborious and lacks scalability. As an alternative, many open source software (OSS) projects employ automated processes for this task, yet this method often relies on large datasets for adequate training. Traditionally, machine learning techniques have been used for issue classification. More recently, large language models (LLMs) have emerged as powerful tools for addressing a range of software engineering challenges, including code and test generation, mapping new requirements to legacy software endpoints, and conducting code reviews. The following research investigates an automated approach to issue classification based on LLMs. By leveraging the capabilities of such models, we aim to develop a robust system for prioritizing issue reports, mitigating the necessity for extensive training data while also maintaining reliability in classification. In our research, we developed an LLM-based approach for accurately labeling issues by selecting two of the most prominent large language models. We then compared their performance across multiple datasets. Our findings show that GPT-4o achieved the best results in classifying issues from the NLBSE 2024 competition. Moreover, GPT-4o outperformed DeepSeek R1, achieving an F1 score 20\% higher when both models were trained on the same dataset from the NLBSE 2023 competition, which was ten times larger than the NLBSE 2024 dataset. The fine-tuned GPT-4o model attained an average F1 score of 80.7\%, while the fine-tuned DeepSeek R1 model achieved 59.33\%. Increasing the dataset size did not improve the F1 score, reducing the dependence on massive datasets for building an efficient solution to issue classification. Notably, some of our models predicted issue labels in individual repositories with precision greater than 98\%, recall of 97\%, and an F1 score of 90\%.

\end{abstract}


\end{frontmatter}


\section{Introduction}
\label{sec:introduction}

Becoming involved in an established OSS project can be a challenge for a developer. Newcomers face several barriers \cite{10.1145/2675133.2675215} when onboarding, and many give up \cite{steinmacher2018almost,constantinou2017empirical,BogdanDisengagement}. The high rate of early dropouts leads to lost opportunities for cultivating the technology workforce and attracting and retaining contributors to projects~\cite{steinmacher2013newcomers}.

Previous research ~\cite{wang2011bug,steinmacher2015understanding,steinmacher2015systematic,stanik2018simple} indicates that newcomers, in particular, face challenges in identifying suitable tasks.

The onboarding of newcomers is essential to keep open OSS projects sustainable~\cite{steinmacher2015systematic}. One of the initial steps of the onboarding process in an OSS project is to find an appropriate task (e.g., bugs, features, etc.) to work with~\cite{wang2011bug,steinmacher2015understanding}. To facilitate task selection, communities add labels to issues, helping new contributors find tasks that are best suited to their capabilities~\cite{steinmacher2018let,santos2022how}. However, in large projects, issue labeling is time-consuming and adds to the workload of already overburdened maintainers~\cite{9057411}. 

Researchers have proposed various approaches to automatically labeling issue types, helping managers prioritize and allocate resources more effectively. \citet{kallis2019tickettagger, kallis2020tickettagger} used fastText to classify issues as bug, feature, or question. Similarly, \citet{colavito2023few} used SETFIT in the 2023 NLBSE competition to predict issue types. \citet{santos2021can} predicted skills to solve an issue using API domains as a proxy and extended to Social Network Analysis (SNA), improving the predictions \cite{santos2023tell}. 

In this study, we leverage LLM APIs to create a fine-tuned model that classifies issues as bugs, features, or questions. Categorizing issues by their types is relevant for triage purposes when maintainers analyze open issues to decide about priorities, allocation \cite{nascimento2024issue} and motivated a tool competition track in a co-located ICSE (International Conference on Software Engineering) conference: Natural Language-based Software Engineering - NLBSE \footnote{\url{https://nlbse2024.github.io/tools/}}.

An LLM is a sophisticated neural network model trained on extensive datasets, including books, code, articles, and websites. This training enables the model to understand the underlying patterns and relationships in language. Consequently, LLMs can produce coherent content, such as natural language and syntactically correct code \cite{ozkaya2023application}. Moreover, an LLM can yield even better results when fine-tuned. We compared the OpenAI and DeepSeek models, building on our previous study. OpenAI, the innovative company behind the development of ChatGPT — the most well-known of the many LLMs they have created — has been a leader in the field. DeepSeek, on the other hand, has emerged as a lightweight competitor with promising initial results.

We fine-tuned and compared several models, evaluating them on two datasets. Fine-tuning is a critical technique in machine learning, particularly in the field of natural language processing (NLP). It involves specialized training on a pre-trained model to enhance performance on specific tasks. Fine-tuning through APIs is a multi-step process that simulates conversations with the LLM and provides the expected responses. By iterating through these simulated conversations, the model learns the appropriate patterns for answering questions, making it a powerful tool for classification. In our approach, we used the title and body of each issue as part of the prompt, with the correct label serving as the expected response.

Finally, since issue triage is a recurring process performed by maintainers in OSS projects, an automatic triage tool that uses LLMs must have a competitive cost-performance ratio.

In this paper, we investigate the following research questions (RQs):

\textbf{RQ1: To what extent can we predict issue types using fine-tuned models?} 
To answer RQ1, we fine-tuned an OpenAI GPT model. RQ1 reproduces our results from the NLBSE 24 tool competition, where a fine-tuned GPT-3.5-turbo model overcame the competition baseline.   

Overall, we found that pre-processing the issue title and body by fine-tuning a GPT-3.5 model can predict the issue types with a macro average of 82.8\% F1 score (table ~\ref{tab:completemetrics}). This yield exceeded the baseline reported in the competition for which this model was initially developed~\cite{nlbse2024}.  





\textbf{RQ2: How does GPT-4o model compare with the previous results?}



RQ2 investigates whether newer versions of the generative AI model can improve the classification of GitHub issues. To explore this, we tested GPT-4o, a more recent version, and compared it with GPT-3.5. Our results demonstrated improvements across all metrics—precision, recall, and F1 score. Overall, we found that a fine-tuned GPT-4o model predicted issue types with an average precision of 86.18\%, as shown in Table  ~\ref{tab:gpt-4o-finetuned}.

\textbf{RQ3: What is the impact of an extended dataset on the models?}

RQ3 aimed to investigate the sensitivity of the fine-tuning process to dataset size and whether it is possible to improve the performance of fine-tuned GPT-4o models by increasing the size of the training and testing datasets.

We used 30,000 data points from the NLBSE 2023 dataset, 
marking a tenfold increase in data compared to the dataset from the 2024 NLBSE tool competition previously used in this study. Our results showed that fine-tuning GPT-4o with the expanded dataset led to a decrease in the average F1 score, producing an F1 score 80.7\% (table ~\ref{tab:gpt-4o-finetuned-extended-dataset}), prompting the question of whether providing more data during fine-tuning could enhance the model’s performance. Surprisingly, this did not seem to be the case. Instead, we observed a decrease in performance, indicating that the quality of the dataset is perhaps more important than the size of the dataset.

Additionally, we also fine-tuned the GPT-4o-mini with this extended dataset to see if it would produce higher results than what was accomplished with the smaller dataset (GPT-4o F1 score of 85.66\% see table~\ref{tab:gpt-4o-finetuned} for full results), but we ended up with an average F1 score of 80.7\% (GPT-4o vs GPT-4o-mini 80.37\% - table~\ref{tab:gpt-4o-finetuned-extended-dataset}). Surprisingly, the results obtained by GPT-4o-mini were nearly identical to those of GPT-4o, but with only 10\% of the cost. This suggests that GPT-4o-mini may be the most cost-efficient model for fine-tuning purposes.

Finally, we evaluated the performance of the new DeepSeek model, comparing it with GPT-4o by fine-tuning both models. For this, we used the DeepSeek-R1-Distill-Llama-8B model from Hugging Face.

We found that fine-tuning DeepSeek-R1-Distill-Llama-8B for GitHub issue classification using the UnsLoTH framework and  Low-Rank Adaptation (LoRA) adaptation led to moderate performance, with an average F1 score of 59.33\% (table ~\ref{tab:metrics-deepseek-r1-finetuned}).

\textbf{RQ4: How does the cost and performance of LLMs models compare for the issue type classification task?}

RQ4 aimed to investigate the cost/performance of the models. We compared the cost of the testing and training process for fine-tuned models and performance.

We found that the costs of GPT-4o are 10 times higher than those of GPT-4o-mini and DeepSeek, while GPT-4o-mini delivered similar results. This indicates that low-cost, fine-tuned models are viable options for the issue triage application.

Our contributions for the issue type classification task includes (1) comparison between vanilla and fine-tuned LLM approaches; (2) A view of LLM models cost-effectiveness; and (3)
The impact of the dataset size increase on the fine-tuned model performance.

The remainder of this paper is organized as follows. Section 2 discusses related work in the field of automated issue classification and the application of LLMs in software engineering. Section 3 describes the methodology, including data preprocessing, model implementation, and fine-tuning strategies. In Section 4, we present and analyze our experimental results based on the three research questions. Section 5 provides a discussion of our findings, including challenges and potential improvements. Section 6 addresses the threats to validity,  Section 7 outlines directions for future work. Finally, Section 8 concludes the paper with a summary of key contributions and observations.

\section{Related Work}
\label{sec:related}


Organizing issues effectively requires proper labeling, which is essential for describing features and enhancing the understanding and retrieval of software artifacts ~\cite{santos2022how}. To support new contributors, social coding platforms like GitHub\footnote{\url{http://bit.ly/NewToOSS}} encourage projects to label beginner-friendly issues, a practice adopted by communities such as LibreOffice,\footnote{\url{https://wiki.documentfoundation.org/Development/EasyHacks}} KDE,\footnote{\url{https://community.kde.org/KDE/Junior_Jobs}}, and Mozilla\footnote{\url{https://wiki.mozilla.org/Good_first_bug}}. However, manually labeling software artifacts is both challenging and time-consuming ~\cite{9057411}. To address this, various approaches have been proposed to automate labeling for software projects \cite{Izadi2021TopicRF, kallis2020tickettagger, el2020automatic}, dependencies \cite{vargas2015automated} or "good first issues," \citet{huang2021characterizing} but this method is limited, as it primarily classifies issues as bugs and does not help project maintainers to triage issues for prioritization or allocation optimization. 

Other efforts have focused on labeling different categories of software artifacts, such as Stack Overflow questions \cite{xia2013tag, lin2019pattern, uddin2019automatic}. \citet{xia2013tag} suggested tags based on question similarity, while \citet{adnan2025sprint} proposed a tool that systematically finds similar issues, evaluates the issue severity, and suggests code location. \citet{uddin2019automatic} and \citet{lin2019pattern} labeled user opinions on APIs to help developers decide whether to adopt a new API. In contrast, \citet{santos2021can, santos2023tell, santos2023tag} focused on supporting task selection based on the APIs that developers encounter in a solution. Knowing in advance the APIs present in a possible solution helps facilitate the newcomer onboarding process. \citet{DBLP:conf/msr/VargovichSPGS23} implemented an API labeling approach using BERT for registered projects, while \citet{carter2025skillscope} analyzed abstract syntax trees (AST) to generate multilevel API labels. The tool leveraged LLMs and Random Forest to predict up to 217 labels. Izadi et al. \cite{izadi2023semantically} created a knowledge graph (KG) that can tag projects enriched by semantic relationships among their topics or tag them using textual information and the existing KG. Even though the aforementioned studies support selecting issues for a contribution, they lack information on the issue's nature or type (e.g., bug or feature), which may be necessary for triage.

To specifically address contribution lifecycle management, projects hosted by Issue Tracking Systems (ITS) utilize labels to streamline the triage ~\cite{nascimento2024issue}. These labels assist maintainers in managing the crucial task of assessing and assigning issues to contributors ~\cite{anvik2006should, anvik2011reducing}. 

The goal of assisting maintainers in handling issues has inspired numerous studies on how to accelerate the triage process. There are several approaches for labeling issues, most focus on distinguishing bug reports from non-bug reports \citep{antoniol2008bug, pingclasai2013classifying, zhou2016combining, zhu2019bug, el2020automatic, perez2021bug}. Later, methods classify issues beyond bug/non-bug reports \citep{kallis2019ticket, kallis2020tickettagger, izadi2022predicting, WANG2021107476} including bug/feature/question labels.

Transformers and LLMs have been applied increasingly to software engineering problems. Notably, \citet{izadi2022predicting} and \citet{WANG2021107476} utilized the BERT text classification algorithm \citep{devlin2019bert} for multi-label classification. \citet{colavito2023few} applied few-shot learning and sentence-BERT (SBERT) for issue tagging. Lin et al. \cite{lin2021traceability} evaluated three BERT architectures for linking artifacts and source code in bug triaging. Beyond issue classification, \citet{liu2020multi} employed a pre-trained transformer-based model for code understanding and generation, while \citet{ahmad2023towards} explored service-based software using AI-powered bots. \citet{santos2023tag} leveraged FastBert to predict skills to solve issues based on categories of APIs. LLMs have also been used in documentation \cite{ozkaya2023application} and code automation \cite{nathalia2023artificial}. 
These studies motivated new research to explore and assess the potential of LLMs to classify issue types \cite{aracena2024applying}.

Colavito et al. \cite{colavito2024leveraging}, leveraged LLMs to predict issue types using a 400-row dataset with zero-shot, few-shot GPT-3.5, and fine-tuned SETFIT, concluding that the negligible difference in favor of SETFIT may not justify the additional cost, given the effectiveness of GPT-3.5 with a zero-shot strategy. In contrast, we explored various low-cost LLM options, including both vanilla and fine-tuned versions, across two dataset sizes. The same authors confirmed that data quality may influence \cite{colavito2024impact} on the prediction results. They filtered out samples based on project stars and project age, and balanced the dataset. The results showed a limited impact from using statistical data quality procedures, and they recommended manual evaluation despite its associated cost. In contrast, we applied cleaning methods to remove noise from issue titles and bodies that could impact text context, following the approach of Santos et al. \cite{santos2023tag}.

In contrast with the aforementioned related work, our study employed a diverse dataset sizes, cleaning methods, LLM APIs and LLM methods to predict issue types, extending our previous study \cite{aracena2024applying}, which outperformed the tool competition baseline \cite{nlbse2024}, with the aim of finding a low-cost and high-performance solution.

\section{Method}

\begin{figure}[!hbt]
\centering
\includegraphics[width=1\textwidth]{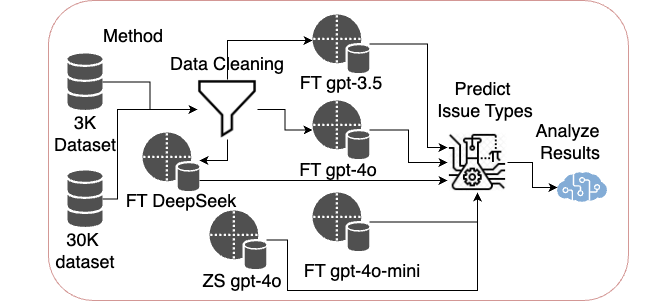}
\caption{Method. Fine-tuned (FT) models and Zero-shot (ZS) Models.}
\label{fig:method}
\end{figure}

Figure \ref{fig:method} depicts our method, composed of data preprocessing, model implementation and training, model evaluation, and analysis. 

\subsection{Data Preprocessing and Cleaning}

\subsubsection{Introduction to the Dataset}

To answer RQ1 and RQ2, we used the NLBSE'24 Tool Competition on Issue Report Classification dataset \cite{nlbse2024}, which is a collection of 3,000 labeled issue reports extracted from five open-source projects. The data were collected over 21 months, from January 2022 to September 2023, covering a wide range of issues. The data were evenly split into a training dataset and a testing dataset. Additionally, the data were distributed evenly across five open-source projects, resulting in 300 data points for each repository in both the training and testing datasets.

To answer RQ3, we expanded the dataset to 30,000 data points, greatly exceeding the size of the NLBSE'24 Tool Competition on Issue Report Classification dataset \cite{nlbse2024}. This dataset was the first 30000 issues from the testing and training dataset of NLBSE 2023 which counted with over 1.4M total issues in its dataset \citet{nlbse2023}

\subsubsection{Attributes of Each Issue Report}
Four key attributes characterize each issue report in the dataset: 

\begin{enumerate}
\item \textbf{Repository}: The name of the open-source project from which the issue report was extracted. This information is only included in the NLBSE 2024 data, which is why individual project results are not available when evaluating NLBSE 2023 models. 
\item \textbf{Label}: The category that the issue report falls into among bug, feature, and question. NLBSE 2023 also included a label called documentation, but since it is not included in the NLBSE 2024 dataset, we excluded it from training and testing.
\item \textbf{Title}: The title of the issue report.
\item \textbf{Body}: The main content of the issue report.
\item \textbf{Author}: This attribute, present only in the NLBSE 2023 dataset but not in the NLBSE 2024 dataset, is intended to identify the author of the issue. However, to ensure consistency when comparing results across both datasets, we kept the same columns. Furthermore, since this attribute often contained “NONE” as its value, we chose to exclude it from our analysis.

\end{enumerate}

\subsubsection{Noise Removal and Data Preparation}
\label{sec:datacleaning}
%
Our research employed two cleaning methods. Although the second method was designed to improve upon the first, we found that some repositories performed better with the initial approach. Consequently, we chose to apply the first method to those repositories. Future research could focus on further optimizing these cleaning methods.

This variance suggests that different data structures, text formats, and content types may respond differently to specific cleaning techniques. Because of this, we selectively applied the first method to repositories where it was more effective. In the following, we provide a detailed comparison of the two methods, highlighting their differences, potential use cases, and why one might be preferred over the other in specific scenarios.

\textbf{Method 1: General Text Standardization and Noise Reduction}

The first method focuses on standardizing textual data and removing non-essential elements to enhance clarity and uniformity. Key steps include:
\begin{enumerate}
\item \textbf{Standardization}: Converting text to lowercase for consistency.
\item \textbf{Noise Removal}: Eliminating double quotes, repository-specific strings (e.g., "DevTools... (automated)"), emojis, URLs, and HTML tags to reduce irrelevant information.
\item \textbf{Word Length Constraint}: Removing words exceeding 20 characters, as research by Miller et al. \cite{miller1958length} indicates, such words make up less than 1\% of English vocabulary and are likely to be accidental noise in the data.
\item \textbf{Whitespace Normalization}: Reducing consecutive whitespaces to a single space to maintain readability and uniformity.
\end{enumerate}

\textbf{Method 2: Enhanced Contextual Cleaning with Structured Transformations}

The second method builds upon the first while incorporating additional refinements to better structure the textual data for downstream analysis. Key differences include:
\begin{enumerate}
\item \textbf{Repository-Specific Adjustments}: Removing additional repository-specific phrases such as "Website or app" and "local React development," which were frequently encountered in our dataset.
\item \textbf{Text Transformation}: Instead of removing all URLs, HTML tags, and user mentions, this method replaces them with standardized placeholders (e.g., \textless URL\textgreater, \textless HTML\_TAG\textgreater, \textless USER\textgreater, \textless IMAGE\textgreater). This preserves contextual indicators for downstream NLP models while eliminating unnecessary noise.
\item \textbf{Markdown Processing}: Stripping unnecessary Markdown syntax such as emphasis markers (*, \_), inline code (`), headers, and links while retaining meaningful text.
\end{enumerate}

\subsubsection{Data Segmentation and Labeling}
The NLBSE 2024 dataset (RQ1 and RQ2) was divided into five repositories, each containing 300 categorized issue reports. The NLBSE 2023 dataset (RQ3) was split into three subsets: one for bugs, one for features, and one for questions (documentation labels were excluded). To maintain the ratio, we selected the first 10,000 issues from each subset and evenly divided them, using half for training and half for testing, while preserving the train-test split ratio from the NLBSE 2024 dataset.

\subsubsection{Format Conversion for Model Input}
\label{sec:format}
When formatting the data for fine-tuning OpenAI's GPT models, training data frames were converted to JSON line files structured as conversations and sent to the API. Each file includes a prompt (\texttt{user\_message}): \textit{"Classify, IN ONLY 1 WORD, the following GitHub issue as "feature", "bug", or "question" based on its title and body}:". The second part (\texttt{assistant\_message}) contains the classification ("bug," "feature," or "question"). Both halves are concatenated and assigned to (\texttt{conversation\_message}),  which is iteratively appended to a single JSON line file.

Meanwhile, when fine-tuning the DeepSeek-R1 model, a structured prompt format (the \textit{train-prompt-style}) with explicit Chain of Thought (CoT) \footnote{as per DeepSeek API guide: https://api-docs.deepseek.com/guides/reasoning\_model} reasoning was used to improve classification accuracy. Unlike the API-based fine-tuning approach used for GPT-4o, which involved direct JSON-formatted conversation data with minimal pre-processing, DeepSeek-R1 required manual prompt engineering and LoRA adaptation to refine classification performance. The train-prompt-style was broken into 3 parts. The first part was the (\texttt{\#\#\# Instruction}): \textit{"You are a GitHub issue classifier that will classify GitHub issues based on their title and description as 'bug', 'feature', or 'question'.
'bug', 'feature', or 'question' are the labels. And these labels are the only possible response.
Please classify the following GitHub issue."}, the second part was (\texttt{\#\#\# Question}): \textit{"issue['title'] + issue['description']"} and the third part was (\texttt{\#\#\# Response}): \textit{"issue['label']"}.

Similar to the GPT approach, we simulated a conversation by providing the expected response, but with additional context and reasoning incorporated into the prompt for the DeepSeek-R1 model.

\subsection{Model Implementation and Training}
\subsubsection{Invoking the API}
Our code utilizes APIs that offer fine-tuning capabilities for LLMs, specifically the GPT-3.5-turbo, GPT-4o, GPT-4o-mini, and DeepSeek R1 models, all of which are renowned for their advanced natural language understanding and generative capabilities. The GPT models can be accessed through OpenAI’s Python library, while the DeepSeek-R1-Distill-Llama-8B model is accessible via the Hugging Face API. These APIs facilitate interaction with language models, enabling the comprehension of prompts and the generation of coherent, contextually appropriate responses.

First, we invoke the OpenAI API. Next, the training files are uploaded to the server. Using these files, fine-tuned models—GPT-3.5-turbo, GPT-4o, and GPT-4o-mini—are created for each repository. To test the model, the interaction with the API occurs through the \texttt{create()}\footnote{openai.chat.completions.create()} method, where a user prompt is constructed. Here, we reuse the prompt from the training cycle, append the testing data, and pass it to the API. The constructed prompt returns its classification of the issue. The API invocation involves parameters such as the model to use, the maximum number of tokens to generate, and the temperature to control the randomness of responses. The model utilized was the fine-tuned model corresponding to its testing dataset, and the maximum number of tokens to be returned was set to 1. The strings "feature," "bug," and "label" were all verified by the OpenAI Tokenizer \footnote{\url{https://platform.openai.com/tokenizer}} to have 1 token each. The temperature was set to 0.0 to minimize randomness in the answers, providing consistent results.

 The process of utilizing the DeepSeek R1 model follows a similar approach to that of the OpenAI models; however, instead of leveraging the OpenAI API, we employed the Hugging Face API. To achieve this, we first authenticated with a Hugging Face token and then used the UnsLoTH API to load the model. From the unsloth API \footnote{\url{https://unsloth.ai/}} , we imported the FastLanguageModel library and utilized the \textit{FastLanguageModel.from\_pretrained} to invoke the \textit{deepseek-ai/DeepSeek-R1-Distill-Llama-8B} model. This approach facilitated seamless interaction with the model, enabling it to be fine-tuned.

\subsubsection{Model Fine-tuning Process (OpenAI Models)}
Fine-tuning a model through OpenAI's API involves a systematic multi-step procedure that requires a comprehensive grasp of the API documentation. As previously outlined, the fine-tuning process was tailored for each repository, demanding dedicated models for enhanced performance.

As explained in section \ref{sec:format}, we created JSON line files customized for individual repositories. These files were uploaded to OpenAI's server to serve as fine-tuning training data. To start the fine-tuning process, we initiated a job to refine the base model using the specified training file. Each job was queued in the cloud environment. Each model required approximately five hours to complete, though the duration varied depending on the number of epochs used. Once all jobs were finished, the fine-tuned models were ready for use.

Upon completion, each fine-tuned model received an ID constructed based on the base model, suffix, prefix, and randomly generated characters. For instance, the model tailored for the "facebook/react" repository was assigned an ID \footnote{(e.g. \textit{ft:gpt-3.5-turbo-0613:gcucst440:fb-issueclassifier:8LLGMnAI})}.

We initially used OpenAI's default hyperparameters for fine-tuning, which automatically configure the learning rate multiplier, the number of epochs, and the batch size. All models started with the standard three epochs, but our training metrics indicated potential for improvement. We then conducted a grid search to optimize the epoch count for each model's specific performance needs.

We retained OpenAI's other default hyperparameters, which are automatically optimized based on dataset characteristics \footnote{https://platform.openai.com/docs/guides/fine-tuning}. However, it is worth noting that both the learning rate multiplier and the batch size can be manually adjusted if needed.

In our experiments, the automatic batch size consistently defaulted to one, although it could be manually set between 1 and 32. The automatic learning rate multiplier remained fixed at two across all experiments, with manual options ranging from 0.1 to 10. According to OpenAI’s documentation, a lower learning rate “may be useful to avoid overfitting,” while batch size influences both the frequency and variance of model parameter updates.

\subsubsection{Model Fine-tuning Process (DeepSeek Model)}
\label{sec:otherLLMs}

The model was initialized using UnsLoTH's FastLanguageModel module, allowing for memory-efficient fine-tuning through 4-bit quantization. The base DeepSeek-R1-Distill-Llama-8B model was loaded with a maximum sequence length of 2048 tokens, optimizing computational efficiency.
To efficiently fine-tune the model, we utilized LoRA \footnote{\url{https://huggingface.co/docs/diffusers/en/training/lora}}, which reduces memory overhead by fine-tuning a subset of parameters. The LoRA configuration targeted key projection layers, utilizing rank-16 adaptation with an alpha scaling of 16.

Fine-tuning was performed using Hugging Face’s SFTTrainer, incorporating gradient accumulation and mixed-precision training to optimize resource utilization.


\subsection{Performance Evaluation Metrics}

In evaluating the effectiveness of our fine-tuned models for issue classification, we utilized three key metrics: precision, recall, and F1 score, as recommended by the competition guidelines
 \cite{nlbse2024}. 

\textbf{Precision}: Precision gauges the accuracy of the positive predictions made by the model. It is the ratio of correctly predicted positive observations to the total predicted positives. High precision indicates that the model is effective in minimizing false positives, which is crucial in scenarios where the cost of a false positive is high.
    
\textbf{Recall}: Recall assesses the model's ability to identify all relevant instances within a dataset. It is the ratio of correctly predicted positive observations to all observations. This metric is vital when the cost of missing a positive prediction, a false negative, is significant.
    
\textbf{F1 score}:  F1 score is the harmonic mean of precision and recall, providing a single metric that balances the two. Maintaining this balance is crucial because prioritizing only precision or only recall can lead to misleading results: high precision with low recall may miss many relevant instances, while high recall with low precision may introduce numerous incorrect classifications.

\begin{equation*}\tag{\roman{num_eq}}
   Precision = \frac{TP}{TP + FP}
\end{equation*}\stepcounter{num_eq}

\begin{equation*}\tag{\roman{num_eq}}
   Recall = \frac{TP}{TP + FN}
\end{equation*}\stepcounter{num_eq}

\begin{equation*}\tag{\roman{num_eq}}
   F1 score = \frac{2TP}{2TP + FP + FN}
\end{equation*}\stepcounter{num_eq}

\subsection{Obtaining the results}
We generated individual CSV files containing the confusion matrix and metrics for each model. Following the evaluation of the five models, we compiled these metrics into a combined CSV file, consolidating the values for True Positives (TP), True Negatives (TN), False Positives (FP), and False Negatives (FN).

The results of each model were gathered into individual and combined CSV files that included confusion matrices and performance metrics. This compilation provided a comprehensive view of the performance of the model, highlighting both areas of strength and those requiring improvement.

Our study calculated these metrics for all models by comparing the predicted model labels against the ground truth from the testing dataset. To ensure reliability and correctness across all models, we utilized the \textit{sklearn.metrics}\footnote{\url{https://scikit-learn.org/stable/api/sklearn.metrics.html}} library to calculate them. The precision, recall, and F1 score collectively offer a holistic view of the model's performance, enabling us to assess its accuracy, robustness, and reliability in classifying issues across different repositories.

We computed metrics for all repositories, and then calculated their average by each metric and label, as shown in table~\ref{tab:completemetrics} and table~\ref{tab:gpt-4o-finetuned}. By combining these results and averaging them, we obtained the overall metrics for each label that can be seen in the table ~\ref{tab:comparisontable}

\section{Results}\label{sec:results}

\paragraph{RQ1 (To what extent can we predict issue types using fine-tuned models?)} 


Table~\ref{tab:completemetrics} shows the average F1 score results when fine-tuning GPT-3.5 model ranged from 76.65\% to 87.08\%. This can be explained by the differences in how the issues were written and the particularities of each repository. Overall, when using \textsc{title} and \textsc{body} combined to fine-tune GPT-3.5, model we reached an overall precision of 83.24\%, recall of 82.87\%, and an F1 score of 82.80\% see table~\ref{tab:completemetrics}.

As seen in \ref{sec:datacleaning}, two different cleaning methods were applied in the models. Initially, we trained the models and obtained metrics for all projects using Method 1. After identifying potential for improvement, we developed and implemented Method 2. However, two of the five models did not exhibit better performance with Method 2, leading us to retain Method 1 for those cases. Further study and analysis are necessary to better understand the underlying reasons for this outcome.

\begin{table}[ht]
\centering
\caption{Fine-tuned GPT-3.5 turbo model metrics by repository and by label}
\small
CM = Cleaning Method ; E = Total Epochs ; P = Precision ; R = Recall
\label{tab:completemetrics}
\begin{tabular}{l|l|l|l|l|l|l}
\hline
\textbf{Repo} & \textbf{CM} & \textbf{E}  & \textbf{Label} & \textbf{P} & \textbf{R} & \textbf{F1}    \\ \hline
facebook   & 1 & 3     & bug            & 0.8333             & \textbf{0.9500}            & 0.8878          \\
facebook   & 1 & 3     & feature        & 0.8557             & 0.8900            & 0.8725          \\
facebook   & 1 & 3     & question       & 0.9024             & 0.7400            & 0.8132          \\ \hline
facebook   & 1 & 3    & average        & 0.8635               & 0.8600            & 0.8579   \\
\hline
tensorflow & 2 & 10 & bug            & 0.9072             & 0.8800            & \textbf{0.8934}          \\
tensorflow & 2 & 10 & feature        & \textbf{0.9318}             & 0.8200            & 0.8723          \\
tensorflow & 2 & 10 & question      & 0.7913             & 0.9100            & 0.8465          \\ \hline
tensorflow & 2 & 10 & average& \textbf{0.8768}             & \textbf{0.8700}            & \textbf{0.8708} \\
\hline
microsoft & 1 & 6      & bug            & 0.8511             & 0.8000            & 0.8247          \\
microsoft & 1 & 6       & feature        & 0.8131                & 0.8700            & 0.8406          \\
microsoft & 1 & 6      & question       & 0.7980             & 0.7900             & 0.7938          \\ \hline
microsoft & 1 & 6      & average& 0.8207             & 0.8200             & 0.8198 \\
\hline
bitcoin & 1 & 3        & bug            & 0.7339             & 0.8000             & 0.7656          \\
bitcoin & 1 & 3        & feature        & 0.8318             & 0.8900            & 0.8599          \\
bitcoin & 1 & 3        & question       & 0.7381             & 0.6200            & 0.6739          \\ \hline
bitcoin & 1 & 3       & average        & 0.7679             & 0.7700            & 0.7665 \\
\hline
opencv & 2 & 6          & bug            & 0.7288             & 0.8600            & 0.7890          \\
opencv & 2 & 6        & feature        & 0.9091             & 0.8000             & 0.8511          \\
opencv & 2 & 6         & question       & 0.8617             & 0.8100            & 0.8351          \\ \hline
opencv & 2 & 6         & average        & 0.8332             & 0.8233            & 0.8250 \\
\hline \hline
all & -- & -- & average & 0.8324 & 0.8287 & 0.8280 \\
\hline \hline
\end{tabular}
\end{table}




\textbf{\emph{RQ1 Summary.}} By reading table ~\ref{tab:completemetrics} we see that it is possible to predict the issue labels with the precision of 83.24\%, recall of 82.87\%, and F1 of 82.8\% using fine-tuned GPT-3.5-turbo base models, with \textsc{title} and \textsc{body} as parameters \footnote{Check the full repository at \url{https://github.com/G4BE-334/NLBSE-issue-report-classification}}.

\paragraph{RQ2: How does GPT-4o  model compare
with the previous results?}

To address RQ2, we retained the cleaning method that yielded the best results in RQ1, and used the same dataset together with the exact fine-tuning process. The base model used was switched from GPT-3.5 to GPT-4o (gpt-4o-2024-08-06), the newest model that could be fine-tuned at that time. As expected, the fine-tuned GPT-4o model outperformed its predecessor across all evaluated metrics, as shown in the comparison table ~\ref{tab:comparisontable}.

\begin{table}[]
\caption{Comparison table NLBSE 2024 baseline model vs GPT-3.5 fine-tuned model vs GPT-4o fine-tuned model}
\label{tab:comparisontable}
\begin{tabular}{l|l|l|l|l}
\hline
\multicolumn{1}{c|}{\textbf{Model}}       & \textbf{Label}                  & \textbf{Precision}                     & \textbf{Recall}                        & \textbf{F1}                            \\ \hline
\multicolumn{1}{c|}{NLBSE 2024 Baseline}  & bug                             & 0.8464                                 & 0.8400                                 & 0.8426                                 \\
\multicolumn{1}{c|}{NLBSE 2024 Baseline}  & feature                         & 0.8448                                 & 0.8700                                 & 0.8426                                 \\
\multicolumn{1}{c|}{NLBSE 2024 Baseline}  & question                        & 0.8001                                 & 0.7700                                 & 0.7827                                 \\ \hline
\multicolumn{1}{c|}{NLBSE 2024 Baseline}  & average                         & 0.8305                                 & 0.8267                                 & 0.8270                                 \\ \hline
{ GPT-3.5 Fine-tuned} & { bug}      & { 0.8109}          & { 0.8580}          & { 0.8321}          \\
{ GPT-3.5 Fine-tuned} & { feature}  & { 0.8683}          & { 0.8540}          & { 0.8593}          \\
{ GPT-3.5 Fine-tuned} & { question} & { 0.8183}          & { 0.7740}          & { 0.7925}          \\ \hline
{ GPT-3.5 Fine-tuned} & { average}  & { 0.8324}          & { 0.8287}          & { 0.8280}          \\ \hline
{ GPT-4.0 Fine-tuned} & { bug}      & { \textbf{0.8538}} & { \textbf{0.8800}} & { \textbf{0.8654}} \\
{ GPT-4.0 Fine-tuned} & { feature}  & { \textbf{0.8843}} & { \textbf{0.8780}} & { \textbf{0.8794}} \\
{ GPT-4.0 Fine-tuned} & { question} & { \textbf{0.8472}} & { \textbf{0.8120}} & { \textbf{0.8249}} \\ \hline
{ GPT-4.0 Fine-tuned} & { average}  & { \textbf{0.8618}} & { \textbf{0.8567}} & { \textbf{0.8566}} \\ \hline
\end{tabular}
\end{table}

\begin{table}[]
\centering
\caption{Comparison table GPT-4o  model vs GPT-4o fine-tuned model with NLBSE 2024 dataset}
\label{tab:gpt-4o-table}
\begin{tabular}{l|llll}
\hline
\textbf{Model}             & \textbf{Repo}       & \textbf{Precision} & \textbf{Recall} & \textbf{F1}     \\ \hline
GPT-4o           & facebook   & \textbf{0.7991}    & \textbf{0.7367} & \textbf{0.7280} \\
GPT-4o           & tensorflow & 0.7261    & 0.6833 & 0.6718 \\
GPT-4o           & microsoft  & 0.7084    & 0.6200 & 0.5448 \\
GPT-4o           & bitcoin    & 0.7558    & 0.7067 & 0.6722 \\
GPT-4o           & opencv     & 0.7616    & 0.6833 & 0.6568 \\ \hline
GPT-4o           & overall    & 0.7502    & 06860 & 0.6547 \\ \hline
GPT-4o fine-tuned & facebook   & 0.8848    & 0.8767 & 0.8739 \\
GPT-4o fine-tuned & tensorflow & \textbf{0.8993}    & \textbf{0.8867} & \textbf{0.8884} \\
GPT-4o fine-tuned & microsoft  & 0.8656    & 0.8633 & 0.8636 \\
GPT-4o fine-tuned & bitcoin    & 0.8001    & 0.8000 & 0.8000 \\
GPT-4o fine-tuned & opencv     & 0.8594    & 0.8567 & 0.8570 \\ \hline
GPT-4o fine-tuned & overall     & 0.8618    & 0.8567 & \textbf{0.8566} \\ \hline
\end{tabular}
\end{table}

\begin{table}[]
\centering
\caption{Fine-tuned GPT-4o model complete metrics table with NLBSE-2024 dataset}
\label{tab:gpt-4o-finetuned}
\begin{tabular}{l|l|lll}
\hline
\textbf{Repo} & \textbf{Label} & \textbf{Precision}          & \textbf{Recall}             & \textbf{F1} \\ \hline
facebook      & bug            & \multicolumn{1}{l|}{0.8362} & \multicolumn{1}{l|}{\textbf{0.9700}} & 0.8981      \\
facebook      & feature        & \multicolumn{1}{l|}{0.8692} & \multicolumn{1}{l|}{0.9300} & 0.8986      \\
facebook      & question       & \multicolumn{1}{l|}{0.9481} & \multicolumn{1}{l|}{0.7300} & 0.8246      \\ \hline
facebook      & average        & \multicolumn{1}{l|}{0.8845} & \multicolumn{1}{l|}{0.8767} & 0.8739      \\ \hline
tensorflow    & bug            & \multicolumn{1}{l|}{0.9263} & \multicolumn{1}{l|}{0.8800} & 0.9026      \\
tensorflow    & feature        & \multicolumn{1}{l|}{\textbf{0.9882}} & \multicolumn{1}{l|}{0.8400} & \textbf{0.9081}      \\
tensorflow    & question       & \multicolumn{1}{l|}{0.7833} & \multicolumn{1}{l|}{0.9400} & 0.8545      \\ \hline
tensorflow    & average        & \multicolumn{1}{l|}{\textbf{0.8993}} & \multicolumn{1}{l|}{\textbf{0.8867}} & \textbf{0.8884}      \\ \hline
microsoft     & bug            & \multicolumn{1}{l|}{0.9053} & \multicolumn{1}{l|}{0.8600} & 0.8821      \\
microsoft     & feature        & \multicolumn{1}{l|}{0.8165} & \multicolumn{1}{l|}{0.8900} & 0.8517      \\
microsoft     & question       & \multicolumn{1}{l|}{0.8750} & \multicolumn{1}{l|}{0.8400} & 0.8571      \\ \hline
microsoft     & average        & \multicolumn{1}{l|}{0.8656} & \multicolumn{1}{l|}{0.8633} & 0.8636      \\ \hline
bitcoin       & bug            & \multicolumn{1}{l|}{0.7941} & \multicolumn{1}{l|}{0.8100} & 0.8020      \\
bitcoin       & feature        & \multicolumn{1}{l|}{0.8788} & \multicolumn{1}{l|}{0.8700} & 0.8744      \\
bitcoin       & question       & \multicolumn{1}{l|}{0.7273} & \multicolumn{1}{l|}{0.7200} & 0.7236      \\ \hline
bitcoin       & average        & \multicolumn{1}{l|}{0.8000} & \multicolumn{1}{l|}{0.8000} & 0.8000      \\ \hline
opencv        & bug            & \multicolumn{1}{l|}{0.8073} & \multicolumn{1}{l|}{0.8800} & 0.8421      \\
opencv        & feature        & \multicolumn{1}{l|}{0.8687} & \multicolumn{1}{l|}{0.8600} & 0.8643      \\
opencv        & question       & \multicolumn{1}{l|}{0.9022} & \multicolumn{1}{l|}{0.8300} & 0.8646      \\ \hline
opencv        & average        & \multicolumn{1}{l|}{0.8594} & \multicolumn{1}{l|}{0.8567} & 0.8567      \\ \hline
overall       & bug            & \multicolumn{1}{l|}{0.8538} & \multicolumn{1}{l|}{0.8800} & 0.8654      \\
overall       & feature        & \multicolumn{1}{l|}{0.8843} & \multicolumn{1}{l|}{0.8780} & 0.8794      \\
overall       & question       & \multicolumn{1}{l|}{0.8472} & \multicolumn{1}{l|}{0.8120} & 0.8249      \\ \hline
overall       & average        & \multicolumn{1}{l|}{\textbf{0.8618}} & \multicolumn{1}{l|}{\textbf{0.8567}} & \textbf{0.8566}      \\ \hline
\end{tabular}
\end{table}

Furthermore, we compared the results obtained from the vanilla GPT-4o model and the fine-tuned GPT-4o model to evaluate the effectiveness of the fine-tuning process. This comparison was not conducted in RQ1, as the primary objective at that stage was to surpass the baseline established by the tool competition.

Overall, the fine-tuned model performed significantly better, as can be seen in table ~\ref{tab:gpt-4o-table} each evaluated metric showed an improvement in precision, recall, and F1 score for each repository. For instance, we observed an increase (31.88\%) in the F1 score for the Microsoft repository model. More notably, the overall F1 score rose from 65.47\% in the vanilla model to 85.67\% in the fine-tuned model.

When fine-tuning GPT-4o and comparing it with the results obtained by fine-tuning GPT-3.5 using the same dataset for testing and training, all evaluated metrics showed improvement. Table~\ref{tab:comparisontable} shows that the overall average precision increased from 83.24\% (table \ref{tab:completemetrics}) to 86.18\%, the overall average recall increased from 82.87\% to 85.67\%, and the overall average F1 increased from 82.80\% to 85.66\%—a significant change. 
Table ~\ref{tab:comparisontable} also compares the baseline, GPT-3.5 and GPT-4o models, where the GPT-4o fine-tuned model outperformed the GPT-3.5 fine-tuned model and the baseline model in every metric—precision, recall, and F1-score—all while utilizing the same training and testing dataset. For example, the precision for bug label on the baseline model was 84.64\%, on the GPT-3.5 fine-tuned model was 81.09\%, and the highest was the GPT-4o fine-tuned model which was 85.38\%. The same holds true for every label and every metric. 

In table~\ref{tab:gpt-4o-table}, it is noteworthy that the fine-tuned GPT-4o model achieved a precision of approximately 90\% for the TensorFlow repository, demonstrating remarkable performance. Conversely, the lowest metrics were observed in the Bitcoin repository, where the model attained approximately 80\% across precision, recall, and F1 score.

Table \ref{tab:gpt-4o-finetuned} depicts GPT-4o metrics per label and project. GPT-4o performance in the TensorFlow project reached 98.82\% of precision for the feature labels. The lowest performance was in the Bitcoin project with 72.73\% in precision, 72\% in recall and 72.36\% F1 score.

\textbf{\emph{RQ2 Summary.}} Fine-tuning GPT-4o demonstrated superior performance across all evaluated metrics—precision, recall, and F1 score—when compared to the fine-tuned GPT-3.5 model on the same dataset, as shown in the table~\ref{tab:comparisontable}.

\paragraph{RQ3: What is the impact of an extended dataset on previous
models?}

To answer RQ3, we added a new extended dataset with the best performance model we found in RQ2, and we also compared GPT-4o , GPT-4o-mini, and DeepSeek.  The goal of comparing GPT-4o, GPT-4o-mini, and DeepSeek was to verify the ability of low-cost models to achieve similar metrics while exploring larger datasets.

We then obtained the dataset for the extended fine-tuning processes using 30,000 datapoints  — 15,000 for training and 15,000 for testing — from the NLBSE 2023 to see if the fine-tuning can scale and improve.

The dataset setup adhered to the following conditions:
\begin{itemize}
    \item We kept the 50\% split for training and testing.
    \item The balance among the issue types was preserved, ensuring an equal number of bug, feature, and question issues.
    \item Each label consisted of 5,000 issues for training and 5,000 issues for testing.
\end{itemize}

\begin{table}[]
\centering
\caption{Comparison table GPT-4o vs GPT-4o-mini fine-tuned models with extended dataset (NLBSE 2023 - 30K)}
\label{tab:gpt-4o-finetuned-extended-dataset}
\begin{tabular}{l|l|l|l|l}
\hline
\textbf{Model}            & \textbf{Label} & \textbf{Precision}          & \textbf{Recall}             & \textbf{F1} \\ \hline
GPT\_4o\_30kDataset       & bug            & \multicolumn{1}{l|}{0.7914} & \multicolumn{1}{l|}{0.8598} & \textbf{0.8242}      \\
GPT\_4o\_30kDataset       & feature        & \multicolumn{1}{l|}{0.8743} & \multicolumn{1}{l|}{0.7748} & 0.8215      \\
GPT\_4o\_30kDataset       & question       & \multicolumn{1}{l|}{\textbf{0.7652}} & \multicolumn{1}{l|}{0.7858} & 0.7753      \\ \hline
GPT\_4o\_30kDataset       & average        & \multicolumn{1}{l|}{0.8102} & \multicolumn{1}{l|}{0.8068} & 0.8070      \\ \hline
GPT\_4o\_mini\_30kDataset & bug            & \multicolumn{1}{l|}{0.7699} & \multicolumn{1}{l|}{\textbf{0.8854}} & 0.8236      \\
GPT\_4o\_mini\_30kDataset & feature        & \multicolumn{1}{l|}{0.8878} & \multicolumn{1}{l|}{0.7564} & 0.8168      \\
GPT\_4o\_mini\_30kDataset & question       & \multicolumn{1}{l|}{0.7713} & \multicolumn{1}{l|}{0.7698} & 0.7706      \\ \hline
GPT\_4o\_mini\_30kDataset & average        & \multicolumn{1}{l|}{0.8097} & \multicolumn{1}{l|}{0.8039} & 0.8037      \\ \hline
\end{tabular}
\end{table}

When evaluating the performance of the fine-tuned GPT-4o model compared with the GPT-4o-mini fine-tuned model, we found that the GPT-4o had a better average F1 score by less than 1\%. The GPT-4o fine-tuned had an average F1 score of 80.7\%, while GPT-4o-mini had an average F1 score of 80.37\% (table ~\ref{tab:gpt-4o-finetuned-extended-dataset}). 

Table ~\ref{tab:metrics-deepseek-r1-finetuned} shows the results of the DeepSeek model. We can observe a decrease in the overall metrics. The highest precision was about 71\%, recall 67\% and F1 score 62\%. The fine-tuned DeepSeek R1 model classifies GitHub issues with moderate success but significant error rates. The confusion matrix (table ~\ref{tab:deepseek-cm}) reveals high true positive counts (3,303 bugs, 2,589 features, 2,869 questions) alongside false positives (2,286 bugs, 1,055 features, 2,678 questions) and false negatives across all categories, indicating that the model needs substantial improvement in distinguishing between issue types.

\begin{table}[]
\centering
\caption{Fine-tuned DeepSeek-R1-Distill-Llama-8B model metrics table with extended dataset}
\label{tab:metrics-deepseek-r1-finetuned}
\begin{tabular}{l|l|l|l|l|}
\hline
\textbf{Model}            & \textbf{Label} & \textbf{Precision} & \textbf{Recall} & \textbf{F1} \\ \hline
DeepSeek\_R1\_fine\_tuned & bug            & 0.5910             & \textbf{0.6744}          & 0.6299      \\
DeepSeek\_R1\_fine\_tuned & feature        & \textbf{0.7105}             & 0.5244          & \textbf{0.6034}      \\
DeepSeek\_R1\_fine\_tuned & question       & 0.5172             & 0.5802          & 0.5469      \\ \hline
DeepSeek\_R1\_fine\_tuned & average        & 0.6062             & 0.5928          & 0.5933      \\ \hline
\end{tabular}
\end{table}



\begin{table}[]
\centering
\caption{Fine-tuned DeepSeek-R1-Distill-Llama-8B model with extended dataset confusion matrix}
\label{tab:deepseek-cm}
\begin{tabular}{l|l|l|l|l|l}
\hline
\textbf{Model}            & \textbf{Label} & \textbf{TP} & \textbf{FP} & \textbf{FN} & \textbf{TN} \\ \hline
DeepSeek\_R1\_fine\_tuned & bug            & 3303        & 2286        & 1595        & 7596        \\
DeepSeek\_R1\_fine\_tuned & feature        & 2589        & 1055        & 2348        & 8788        \\
DeepSeek\_R1\_fine\_tuned & question       & 2869        & 2678        & 2076        & 7157        \\ \hline
\end{tabular}
\end{table}

\textbf{\emph{RQ3 Summary.}} Table ~\ref{tab:gpt-4o-finetuned-extended-dataset} shows our evaluation of  and GPT-4o-mini using a dataset 10 times larger than the one used for the fine-tuned model in table ~\ref{tab:gpt-4o-finetuned}. This expanded evaluation revealed a significant performance decline: the average precision dropped from 86.18\% to 81.02\%, average recall fell from 85.67\% to 80.68\%, and the average F1 score decreased from 85.66\% to 80.70\%.

\paragraph{RQ4: How does the cost and performance of LLMs models compare for the issue type classification task?}

We wanted to know whether the costs of fine-tuning the GPT-4o model are producing a difference. To analyze that, we evaluated the tests on a vanilla version of GPT-4o by calling the API and then compared it with a fine-tuned version of GPT-4o.

Using the vanilla GPT-4o model incurs less than 1\% of the cost of the fine-tuning process. However, our research demonstrates that despite the additional cost, fine-tuning the model led to a precision increase of approximately 27\% for the TensorFlow project and 9\% for Facebook, representing the best performance achieved with both GPT-4o and the fine-tuned GPT-4o model.

Fine-tuning the GPT-4o-mini model costs \$3.00 per 1M tokens, totaling under \$10.00 for testing and training. In contrast, GPT-4o costs \$25.00 per 1M tokens, resulting in a total cost of just over \$100 for testing and training. Based on these results, it can be confidently concluded that GPT-4o-mini is an optimized, cost-efficient version of GPT-4o, offering minimal reductions in performance. DeepSeek costs are cheaper to GPT-4o-mini, \$0.14 per 1M tokens, totaling under \$1.00 for testing and training; however, for the classification task, it failed to obtain a competitive F1 compared with GPT-4o-mini.

\textbf{\emph{RQ4 Summary.}} The costs for fine-tuning model with GPT-4o-mini are 10\% of the costs of GPT-4o while delivers similar results for the issue type classification task.

\section{Discussion}
\label{sec:discussion}

\paragraph{Cleaning Method Performance and Observations}

Our results indicate that Method 1 performs better for repositories characterized by primarily plain text data, where excessive transformations are unnecessary. This method ensures a clean yet minimally altered dataset, making it well-suited for direct text analysis. In contrast, Method 2 proved more effective for repositories containing structured text elements such as Markdown, extensive links, or HTML-based formatting. By replacing rather than removing certain elements, Method 2 preserves contextual meaning that might otherwise be lost during the cleaning process.

The performance discrepancies between the two methods highlight the importance of context-aware text cleaning. While a general cleaning approach may work for some datasets, structured content often requires more nuanced transformations. Future work could explore adaptive cleaning techniques that dynamically adjust based on text characteristics, further optimizing data preprocessing for diverse repositories.

\paragraph{What are the difficulties in labeling?} Ambiguous labeling, especially for the "question" label, poses a significant challenge in issue classification. Observing the results obtained on table~\ref{tab:completemetrics}, it consistently shows that the "question" labels exhibit the lowest precision, recall, and F1 scores across repositories compared to the "bug" and "feature" labels.

To better understand this phenomenon, we conducted a detailed analysis of misclassified examples. For instance, in the Facebook repository, an issue labeled as a question had the title “Bug: useRef can not return a persist ref object,” which suggests that the correct label should have been bug, as it describes incorrect functionality. In another case from the same test dataset, an issue labeled as a question contained only the statement: “language and translation i’m sure that you translated your react site by Google the worst result at all i hope you correct it…,” without any interrogative structure, making the question label inappropriate.

The classification difficulty appears to stem from several factors. First, questions in GitHub issues often lack explicit interrogative markers—many are implicit requests embedded within problem descriptions. Second, users frequently mislabel issues as questions when they could be more accurately categorized as bugs or features. Third, the semantic boundaries between reporting a problem (bug) and seeking information about a problem (question) are often blurred in natural language descriptions.

In a representative example, a user described technical challenges with Redux state persistence after migrating from Django to React. Although structured as a troubleshooting report without explicitly phrased questions, it was labeled as a question—likely because the user was implicitly seeking assistance. Conversely, another issue explicitly asked, “Is this behavior expected?”, making it a clear candidate for the question label. These examples illustrate how questions can be embedded within broader narratives or conveyed without conventional interrogative structures.

For future work, we propose three approaches to improve the classification of ambiguous "question" labels.
The first will structure prompts to guide the language model in identifying both explicit and implicit questions, with special attention to interrogative intent rather than just syntactic markers.
Next, we will incorporate linguistic features that capture interrogative semantics beyond superficial patterns like question marks or interrogative pronouns.
Finally, we will implement a hierarchical approach that first distinguishes between information-seeking (questions) and information-providing (bugs/features) issues before further classification.
These improvements could significantly enhance automated labeling accuracy, particularly for the challenging "question" labeling, while accommodating the diverse linguistic patterns and domain-specific terminology found across different repositories. The techniques may be adapted for different issue types.

Similarly, when comparing the metrics on different repositories, it is clear that the results for the bitcoin repository were worse than those of the other repositories. When we compare the F1 score with the "tensorflow" repository (87.08\%), the F1 score of the bitcoin repository is more than 10\% worse (76.65\%). 
After further analysis, we concluded that it was due to the poor labeling and description of the issues by the developers who work on each repository. We concluded that, upon comparing our results with the baseline metrics, the baseline model showed consistently worse overall performance when classifying issues from the bitcoin repository. Additionally, many of the bitcoin issues were written without much clarity. Like the following issue that was labeled as a question by the user "\textit{Fixed some times can not remove '\$SUFFIX-dirty' on version number correctly with git tag using Gitian. Even if you use a git annotated tag, you encounter such a problem.}"




Looking at the confusion matrix in table \ref{tab:confusionM}, one can see that 
models have varying degrees of success in classifying different types of issues (bugs, features, questions) across the different repositories. The model consistently identifies "bug" labels across repositories, with TP ranging from 80 to 89, suggesting that the features used by the model are good indicators of this class. The models seem to perform well on "feature" classification, with relatively high TP and higher FP than "bug" classification. This could indicate confusion between "feature" and other types of issues, leading to more false alarms. There is a notable variance in the model's ability to correctly classify "question" labels, with the lowest TP observed in the "bitcoin/bitcoin" repository.

The "facebook/react" repository has relatively balanced classification across all labels, indicating that the model may have learned the distinctive features of each label well in this context. On the other hand, the "tensorflow/tensorflow" repository has the highest FP for the "question" classification, suggesting the potential overclassification of this label for this fine-tuned model. Additionally, the "microsoft/vscode" repository tends to miss "feature" and "question" issues (as indicated by higher FN), which might suggest that for the Microsoft VS Code contributors, these labels are too broad and sometimes misleading.


\begin{table}[h!]
\centering
 \begin{center}
 \caption{Fine-tuned GPT-3.5 model with NLBSE 2024 dataset confusion matrix}
 \label{tab:confusionM}
\begin{tabular}{l|l|l|l|l|l}
 \hline
\textbf{Repository}   & \textbf{Label} & \textbf{TP} & \textbf{FP} & \textbf{FN} & \textbf{TN} \\ \hline
facebook/react        & bug            & 89          & 15          & 11          & 185         \\
facebook/react        & feature        & 95          & 19          & 5           & 181         \\
facebook/react        & question       & 74          & 8           & 26          & 192         \\ \hline
tensorflow/tensorflow & bug            & 82          & 6           & 18          & 194         \\
tensorflow/tensorflow & feature        & 88          & 9           & 12          & 191         \\
tensorflow/tensorflow & question       & 91          & 24          & 9          & 176         \\ \hline
microsoft/vscode      & bug            & 87          & 20          & 13          & 180         \\
microsoft/vscode      & feature        & 80          & 14          & 20          & 186         \\
microsoft/vscode      & question       & 79          & 20          & 21          & 180         \\ \hline
bitcoin/bitcoin       & bug            & 89          & 18          & 11          & 182         \\
bitcoin/bitcoin       & feature        & 80          & 29          & 20          & 171         \\
bitcoin/bitcoin       & question       & 62          & 22          & 38          & 178         \\ \hline
opencv/opencv         & bug            & 80          & 8          & 20          & 192         \\
opencv/opencv         & feature        & 86          & 32          & 14          & 168         \\
opencv/opencv         & question       & 81          & 13          & 19          & 187         \\ \hline

\hline

 \end{tabular}
 \vspace{-10pt}
 \end{center}
\end{table}
Our models could likely benefit from additional tuning or training data to improve classification, especially for "question" labels.
Addressing the imbalance between FP and FN across different labels could help improve model performance. This might involve re-evaluating the features used for classification and tuning the learning rate multiplier.

Compared to the baseline results 
our model performed slightly better overall. The baseline model utilized the \textit{all-mpnet-base-v2} sentence transformer developed by Hugging Face as their base model and fine-tuned it with SetFitTrainer. The baseline model, as seen in table~\ref{tab:comparisontable}, had an overall average F1 score of 82.7\% when we got 82.8\%. The average F1 score on the "feature" and "question" labels for the baseline model was 84.26\% and 78.27\%, respectively. Meanwhile, our models produced 85.93\% and 79.25\% average F1 score, respectively.

The fine-tuned GPT model exhibits heterogeneous performance across issue classification categories; as previously noted, issues labeled as questions demonstrate notably lower accuracy compared to those labeled as bugs or features. Analysis of the confusion matrices (table ~\ref{tab:confusionM}) reveals systematic misclassification patterns that likely stem from the model's interpretation of linguistic and contextual cues. The model's decision-making process appears to struggle with distinguishing between implicit questioning and problem statements, particularly in repositories with domain-specific terminology (e.g., bitcoin/bitcoin with 38 false negatives for questions). False positives in question classification (prominently in tensorflow/tensorflow with 24 instances) suggest the model may erroneously interpret uncertainty in technical descriptions as interrogative intent. Conversely, false negatives likely occur when questions are framed within technical contexts that obscure their interrogative nature. Feature classification demonstrates a tendency toward overprediction (exhibited by high false positives in opencv/opencv and bitcoin/bitcoin), indicating potential confusion between aspirational descriptions and actual enhancement requests. These findings suggest that the model's internal representations inadequately capture the nuanced contextual factors that distinguish questions from other issue types across varying technical domains and discourse conventions.

\paragraph{To what extent can the updated versions of GPT model improve the classification of issues?}
Another significant question arising from the conference was whether future versions of generative AI models would enhance the classification of GitHub issues. To investigate this, GPT-4o—an advanced and updated version of ChatGPT compared to GPT-3.5—was evaluated. Improvements were observed across all metrics, including precision, recall, and F1 score. Notably, the enhancements were not limited to overall averages but reflected consistent improvements across each individual repository. This is best displayed in table~\ref{tab:gpt-4o-finetuned}, where we can check the broken-down metrics for every model.

\paragraph{What is the impact of the extended dataset?}
It was observed that the fine-tuned GPT-4o model, when trained on a dataset of 3,000 data points from NLBSE 2024, achieved an overall average F1 score of 85.66\% (table~\ref{tab:comparisontable}). In contrast, fine-tuning the same model with a significantly larger dataset of 30,000 data points from NLBSE 2023 resulted in a lower F1 score of 80.70\% (table~\ref{tab:gpt-4o-finetuned-extended-dataset}).

These findings suggest that increasing the size of the dataset by a factor of ten, for both training and testing, does not necessarily guarantee an improvement in model performance. 
A possible explanation for this phenomenon is that the quality of fine-tuning data is more critical than its quantity, particularly for LLMs such as OpenAI’s GPT models, which have already undergone extensive pre-training on large-scale datasets.

The NLBSE 2024 dataset, used in the initial fine-tuning models, was carefully curated by the NLBSE tool competition team, ensuring the inclusion of highly relevant data points and the exclusion of outlier data points. Conversely, the NLBSE 2023 dataset, despite being larger, was not subject to the same rigorous selection process due to resource constraints. It was a large selection of 1.4M data points, therefore a curating process here would be unfeasible. Additionally, when utilizing this dataset, we also did not have enough resources to handpick the best 30,000. Instead, the first 30,000 available data points were selected, with 15,000 used for training and 15,000 for testing, while maintaining a balanced distribution of 5,000 instances per label (i.e., "bug," "question," and "feature") per process (i.e, training, and testing).  

These results reinforce the importance of data quality over sheer volume in the fine-tuning process, emphasizing that careful dataset selection is essential for optimizing model performance.


A similar trend was previously observed in the NLBSE 2023 tool competition results for issue report classification. LLMs undergo extensive pretraining on vast datasets, far exceeding what individuals or organizations without the resources of major technology companies can provide. Consequently, increasing the dataset size during fine-tuning appears to have minimal, or in some cases even negative, effects on model performance. This is evident in table~\ref{tab:gpt-4o-finetuned-extended-dataset}, where fine-tuning GPT-4o and GPT-4o-mini with a dataset ten times larger than that used in table~\ref{tab:comparisontable} led to a decline in performance. Specifically, 
 the average recall dropped from 85.67\% to 80.68\%, and the F1 score fell from 85.66\% to 80.70\%.

These results reinforce the hypothesis that, when fine-tuning LLMs, the quality of the dataset may be more critical than its size. This finding was also observed by ~\citet{colavito2023few} in their submission for the NLBSE 2023 issue report tool competition, where they demonstrated that manually selecting and approving issues for a few-shot training and testing approach resulted in a 6\% improvement over using the entire dataset.. Using the manually selected dataset, the researchers obtained an average F1 score of 83.21\%. In contrast, they received an average F1 score of 77.67\% when testing for all issues, which shows that by selecting the dataset presented better results on top of being more cost-efficient.

\begin{table}[]
\centering
\caption{Baseline models metrics for NLBSE 2023 tool competition}
\label{tab:nlbse-2023-baseline}
\begin{tabular}{l|l|l|l|l|}
\hline
\textbf{Model} & \textbf{Label} & \textbf{Precision} & \textbf{Recall} & \textbf{F1} \\ \hline
FastText       & bug            & 0.8771             & 0.9173          & 0.8967      \\
FastText       & feature        & 0.8415             & 0.8621          & 0.8517      \\
FastText       & question       & 0.6702             & 0.5011          & 0.5964      \\
\hline
FastText       & average        & 0.8510             & 0.8510          & 0.8510      \\ \hline
RoBERTa        & bug            & \textbf{0.9110 }            & \textbf{0.9390}         & \textbf{0.9248}      \\
RoBERTa        & feature        & 0.8950             & 0.8967          & 0.8958      \\
RoBERTa        & question       & 0.7309             & 0.5684          & 0.6395      \\
\hline
RoBERTa        & average        & 0.8906             & 0.8906          & 0.8906      \\ \hline
\end{tabular}
\end{table}

Despite our model not beating the baseline metrics proposed by the NLBSE 2023 competition (85.1\% F1 score and 89.06\% F1 score - table ~\ref{tab:nlbse-2023-baseline}), the fine-tuned GPT-4o-mini produced an F1 score of 80.37\% (table ~\ref{tab:gpt-4o-finetuned-extended-dataset}) which is higher than the Colavito et al. score 77.67\% F1 score ~\cite{colavito2023few} while being more time and cost efficient.

Compared to the baseline models FastText and RoBERTa, our models were not able to outperform the NLBSE 2023 tool competition metrics. However, the performance when labeling "questions" was overall better in all metrics (precision, recall, and F1 score). As a matter of fact, the precision obtained by the GPT-4o fine-tuned model with the extended dataset was the highest among all of them presented in the paper, with a magnificent 94.8\% (table \ref{tab:gpt-4o-finetuned-extended-dataset}), which means the model performed well on classifying as a question when it is entirely sure it is a question.

\paragraph{What is the difference in performance between fine-tuning models considering their cost?}
The difference is slight, and for cost saving, GPT-4o-mini would be preferable. When considering the cost per token on these fine-tuning API calls, it is clear that fine-tuning a GPT-4o-mini model is significantly more cost-effective.

It is also clear in our evaluation that the fine-tuned GPT-4o model outperformed DeepSeek R1 even when using the same dataset. The fine-tuned DeepSeek R1 model obtained an average F1 score of 59.33\% (see table~\ref{tab:metrics-deepseek-r1-finetuned}), which was significantly worse than the GPT-4o fine-tuned model. 

The DeepSeek R1 model struggled with recall, particularly in distinguishing bugs from questions, and required more resource-intensive fine-tuning compared to GPT-4o. These results highlight the challenges in adapting DeepSeek-R1 for domain-specific classification tasks, suggesting the need for further dataset balancing and optimization of fine-tuning parameters.

Despite not producing great metrics overall, DeepSeek-R1 has been proven to be very cost effective. It is free to fine-tune a DeepSeek-R1 through Hugging Face; the only cost associated with it is to use GPU Power from Google Colab ~\footnote{ https://colab.research.google.com/drive/1z\_86hKTwxL2O4PHsxtT5\_VZONHPUBFPW?usp=sharing} since the fine-tuning process requires GPUs for it to run effectively. A \$10.00 fee was paid to get access to Google Colab Pro and be able to use Google's Colab most powerful GPUs. When compared to the cost of fine-tuning GPT-4o, the cost of fine-tuning GPT-4o-mini is one-tenth, similar to the total cost of DeepSeek-R1 (fine-tuning and GPU). Since fine-tuning GPT-4o-mini produced results comparable to GPT-4o but at a significantly lower cost, it can be considered the most cost-efficient model for fine-tuning.

\section{Threats to Validity}
\label{sec:threatsvalidity}

We divided the threats to validity into construct, external, and internal validity.

\paragraph{Internal Validity} One potential threat is the preprocessing techniques applied to the dataset, which may introduce biases that affect model performance. To mitigate this, we experimented with multiple data-cleaning methods and validated our approach against a benchmark dataset. 
\paragraph{External Validity} Our dataset is limited to issue reports from five repositories, which may not represent all open-source software projects. We utilized available datasets from tool competitions, which consist of repositories from various domains and sizes. However, further studies using more diverse datasets are needed to assess the generalizability of our approach. Data leakage presents a potential threat to the validity of this study, as it may limit the generalizability of the results to real-world applications where sensitive data could be inadvertently exposed by the LLM. 
\paragraph{Construction Validity} A key threat is the correctness of issue labels assigned in the dataset, as misclassified labels could impact model training and evaluation. We mitigated this by relying on an existing labeled dataset from a recognized competition. Additionally, using only three issue types (bug, feature, and question) may not capture the complexity of real-world classification scenarios. Future work should explore more granular categorization. Prompt engineering techniques can change the results. Future work may explore different prompt techniques to evaluate the impact on the results.

\section{Future Work}
While fine-tuning large language models has demonstrated notable improvements in issue classification, our findings suggest that data quality is a more significant factor than dataset size. Given these insights, a promising future research direction involves the integration of Retrieval-Augmented Generation (RAG) to enhance model performance.
RAG combines pre-trained LLMs with an external knowledge retrieval mechanism, enabling models to access and incorporate relevant, real-time information during inference. This approach could mitigate the limitations of static fine-tuning by dynamically incorporating domain-specific context from GitHub repositories, documentation, and historical issue resolutions. 
By leveraging RAG, an LLM could retrieve relevant issues, discussions, and resolutions from a knowledge base before classifying a new issue. This would not only enhance precision but also reduce the need for extensive fine-tuning, making the approach more scalable and adaptable across different software projects. Future work should explore the effectiveness of RAG in improving classification accuracy, particularly for ambiguous labels such as "question," where contextual understanding plays a critical role.

To handle the ambiguous labeling of question labels, we propose developing a two-stage classification pipeline that first distinguishes between information-seeking (questions) and information-providing (bugs/features) issues before proceeding to more granular classification. This approach may better capture the semantic nuances that our current models struggle to differentiate, particularly when issues contain both problem descriptions and implicit requests for information.

Additionally, our findings regarding the comparable performance of GPT-4o-mini to GPT-4o at significantly lower cost suggest an important area for future investigation. Research should systematically evaluate the cost-performance trade-offs across different model sizes, particularly exploring whether smaller, more efficient models can achieve comparable classification accuracy with appropriate fine-tuning strategies.

To enhance the interpretability of our model’s predictions, we plan to incorporate explainable AI (XAI) techniques such as SHAP (Shapley Additive Explanations) and LIME (Local Interpretable Model-agnostic Explanations). These methods will help identify which textual features contribute most to misclassifications, thereby guiding model refinement and dataset augmentation in addition to incorporating human feedback to validate and refine predictions \cite{wang2023decodingtrust} By leveraging active learning frameworks, we can iteratively retrain the model on hard-to-classify cases, reducing the occurrence of false positives and false negatives over time.

By pursuing these research directions, we aim to address the remaining challenges in automated issue classification, particularly for ambiguous categories like "question" labels, while maintaining the efficiency and accessibility advantages demonstrated by our current approach.



\section{Conclusion}

In conclusion, this study applies large language models, specifically OpenAI's GPT-3.5-turbo, GPT-4o-mini, and GPT-4o, as well as DeepSeek's R1 model, to classify GitHub issue reports. By fine-tuning models on datasets from five distinct repositories, we demonstrated the feasibility and efficiency of this approach. Our methodology, focusing on data preprocessing and model fine-tuning, yielded an average F1 score of 85.66\%, ultimately surpassing the baseline model's effectiveness in classifying issues into "bug," "feature," or "question" categories. However, this performance varied across repositories, highlighting the nuanced nature of issue report classification.

If OSS were to utilize a curated large proportion of its issues for fine-tuning, the classification performance would likely improve further, yielding even higher results. The past NLBSE tool competition, for instance, provided a dataset with 1.6 million issues as a reference ~\cite{colavito2023few}, suggesting a potential avenue for further exploration. 

One of the key findings was the variability in performance based on the nature of the data in each repository. This underscores the necessity of tailored approaches when applying language models in diverse contexts. Furthermore, our study identified challenges in classifying "question" labels due to their often ambiguous nature. This points to a broader issue in the standardization of labeling practices within the GitHub community.
Future research could further investigate text-cleaning strategies for different datasets.

Despite fine-tuning efforts, DeepSeek-R1-Distill-Llama-8B underperformed compared to OpenAI's GPT-4o. The model struggled to achieve high recall for feature classification and frequently misclassified issue labels. Additionally, the fine-tuning process for DeepSeek-R1 proved more complex and resource-intensive than OpenAI's GPTs, requiring the use of GPU for fine-tuning, indicating potential inefficiencies in adaptation workflows for certain tasks.
The findings suggest that DeepSeek-R1 has not shown to be a viable alternative for classifying issues. On top of that, OpenAI's fine-tuning process provides a more streamlined experience with superior classification performance.

\section*{Acknowledgements}
\label{}
We would like to thank Dr. Isac Artzi of GCU for partially supporting this work, especially by providing us with the equipment that was used in this research. This work was also supported by the National Science Foundation under Grant Numbers 2236198, 2247929, 2303042.

\bibliographystyle{ACM-Reference-Format}
\bibliography{references}

\newenvironment{myquote}%
  {\list{}{
    \leftmargin=0.5in%
    \rightmargin=0in%
  }\item[]}%
  {\endlist}

\newcommand{\reviewer}[4][R]{\textbf{#1#2.#3:} \itshape #4 \upshape}
\newcommand{\answer}[4][R]{\textbf{Answer to #1#2.#3:}\xspace#4\vspace{4mm}}
\newcommand{\answerquote}[1]{\begin{myquote}\setlength{\parskip}{1.5mm}{\fontsize{9pt}{9pt}\selectfont#1}\end{myquote}}
\newcommand{\revref}[3][R]{\textbf{#1#2.#3}}
\newcommand{\revsection}[2][\normalsize]{\vspace{3mm}\hrulefill\hspace{2mm}{#1\uppercase{#2}}\hspace{2mm}\hrulefill\vspace{3mm}}

\renewcommand{\responsetoreviewer}[3][black]{#3}

\clearpage
\pagestyle{empty}
\setlength{\parindent}{0cm}
\setlength{\parskip}{1.5mm}
\normalsize

\end{document}